# Local stability and evolution of the genetic code

V. R. Chechetkin[a]* and V.V. Lobzin[b]

[a]*Theoretical Department of Division for Perspective Investigations, Troitsk Institute of Innovation and Thermonuclear Investigations (TRINITI), Troitsk, 142190 Moscow Region, Russian Federation*
[b]*School of Physics, University of Sydney, Sydney, NSW 2006, Australia*

**Abstract**

The standard genetic code is known to be robust to translation errors and point mutations. We studied how small modifications of the standard code affect its robustness. The robustness was assessed in terms of a proper stability function, the negative variations of which correspond to a more robust code. The fraction of more robust codes obtained under small modifications appeared to be unexpectedly high, about 0.1–0.4 depending on the choice of stability function and code modifications, yet significantly lower than the corresponding fraction in the random codes (about a half). In this sense the standard code ought to be considered distinctly non-random in accordance with previous observations. The distribution of the negative variations of stability function revealed very abrupt drop beyond one standard deviation, much sharper than for Gaussian distribution or for the random codes with the same number of codons in the sets coding for amino acids or stop-codons. This behavior holds for both the standard code as a whole and its binary NRN-NYN, NWN-NSN, and NMN-NKN blocks. Previously, it has been proved that such binary block structure is necessary for the robustness of a code and is inherent to the standard genetic code. The modifications of the standard code corresponding to more robust coding may be related to the different variants of the code. These effects may also contribute to the rates of replacements of amino acids. The observed features demonstrate the joint impact of random factors and natural selection during evolution of the genetic code.

## 1. Introduction

The origin and evolution of the genetic code remains still the fundamental unsolved problem (a critical summary of state-of-the-art may be found in Barbieri, 2008; Koonin and Novozhilov, 2009). The two main concepts concerning the code origin are based on the physicochemical correspondence between the cognate (anti)codons and amino acids (Gamow, 1954; Woese et al., 1966; Pelc, 1965; Pelc and Welton, 1966; Jungck, 1978; Root-Bernstein, 1982; Blalock and Smith, 1984; Yarus, 1998; Yarus et al., 2005; Copley et al., 2005) and the biosynthetic pathways (Wong, 1975; Szathmary, 1993; Amirnovin, 1997; Di Giulio and Medugno, 1998; Di Giulio, 2005; Wong, 2005). Crick (1968) stressed the important role of random factors during evolution of the genetic code.

Already at the early stage of investigations, it has been recognized that the code should be robust with respect to the effects of possible misreading during translation and to the changes of codons caused by point mutations (Woese, 1965). This concept was laid to the ground of the error-minimization theory (Alff-Steinberger, 1969; Haig and Hurst, 1991; Goldman, 1993; Freeland and Hurst, 1998; Ardell, 1998; Gilis et al., 2001; Luo and Li, 2002; Goodarzi et al., 2005; Sella and Ardell, 2006; Novozhilov et al., 2007). The standard genetic code was proved to be much more efficient in minimization of adverse effects of translation errors and point mutations as compared with a random code of similar block structure. The estimates obtained versus random codes with the same number of codons in the coding sets (Haig and Hurst, 1991; Freeland and Hurst, 1998; Goodarzi et al., 2005) showed that the chance to find a code more robust than the standard one is very small.

Using different cost functions and swaps of codon batches between coding sets as permissible modifications of the genetic code, Novozhilov et al. (2007) studied possible evolutionary trajectories starting from a random code and from the standard one. In the latter case they found rather frequent events leading to the lowering of cost function. Novozhilov et al. (2007) concluded that the difference between the standard code and a partially optimized random code is not so crucial and that the evolution of the code can be represented as a combination of adaptation and frozen accident. These results prove also that the assessment of fraction of codes more robust than the standard one depends significantly on the choice of representative codes (random or slightly modified variants of the standard code) and raise the general problem of local stability of the genetic code under small modifications. In this paper we extended the analysis of local stability to the broader class of modifications beyond the relatively narrow space of the codes with the same number of codons in the coding sets, studied the stability in the separate binary blocks of the standard code, outlined the scheme for the

*Corresponding author. *E-mail addresses:* chechet@biochip.ru ; vladimir_chechet@mail.ru

comparison of local stability of the code variants and suggested the model for simulation of the possible replacements of amino acids under more robust coding.

Choosing proper stability function, we studied how small modifications of the standard code affect its robustness. The negative variations of stability function under such modifications correspond to more robust codes. The fraction of more robust codes obtained under small modifications appeared to be about 0.1–0.4 depending on the choice of stability function and code modifications, i.e. much higher than the estimates obtained versus random codes with the same number of codons in the coding sets (Haig and Hurst, 1991; Freeland and Hurst, 1998; Goodarzi et al., 2005). This value is nevertheless significantly lower than the corresponding fraction in the random codes (about a half) and proves the non-randomness of the standard code. The distribution of negative variations of stability function drops rapidly beyond range of one standard deviation, much faster than for Gaussian distribution or for random codes with the same number of codons in the coding sets. The latter feature was proved by the control simulations for random codes. This behavior holds for both the standard code as a whole and its binary NRN-NYN, NWN-NSN, and NMN-NKN blocks. As has been proved earlier, such binary block structure is necessary for the robustness of the genetic code coding for hydrophobic/hydrophilic amino acids and/or amino acids with large/small molecular volumes (Chechetkin, 2003) and is inherent to the standard code (Rumer, 1968; Jungck, 1978; Wolfenden et al., 1979; Blalock and Smith, 1984; Taylor and Coates, 1989; Chechetkin, 2003, 2006; Wilhelm and Nikolaeva, 2004). It will be shown below that the concomitant binary block structure of the genetic code imposes certain bias on the distribution of negative variations of stability function.

The plan of the paper is as follows. Section 2 describes the general scheme for the assessment of stability of the standard code under small modifications as well as the characteristics referring to the whole code. The stability of binary blocks of the standard code is considered in Section 3. The genetic code should be fixed and became nearly universal at the early stage of molecular evolution (Crick, 1968; Freeland et al., 2000). Could the factors related to the more efficient error-minimization play some role in the present conditions? We discuss the possible relationship of these effects to the replacements of amino acids (Section 4) and to the known variants of the genetic code (Section 5). Our results indicate that small modifications leading to more robust coding may participate in the development of alternative codes. Finally, we discuss the limitations of the approach based on the stability function and the possible consequences from our results. The study of the coding stability under small modifications of the standard code permits to elucidate the possible evolutionary pathways at the origin and evolution of the genetic code.

## 2. Stability of the standard code under small modifications

### 2.1. Modifications

In our scheme the small modifications of the code were produced (i) by picking up a codon from a set coding for a particular amino acid and transposing it to a set coding for a different amino acid and containing a codon close (differing only by one-nucleotide replacement) to initial one and (ii) by the swaps of close codons in different sets. For brevity, the first operation will be termed "delin" (deletion of a codon from one set and insertion it to the other). The delin operation changes the number of codons in the sets coding for the different amino acids and simulates the effects of misreading and point mutations. As an example of this operation let the codon GGA coding for glycine be picked up. Among nine neighbors differing from GGA by one-nucleotide replacements, three (GGC, GGU, and GGG) belong to the same set coding for glycine. The delin operations within the same set retain the standard code intact and are discarded. Six other delin operations lead to the discernible modifications of the standard code and are taken into account for comparison of robustness. The swap of close codons in a pair of sets coding for different amino acids is an elementary operation retaining the total number of codons in each set. We will consider two variants of the code, with and without stop-codons (64 and 61 codons in the code, respectively).

### 2.2. Stability function

The quantitative assessment of the code robustness is performed with a stability function (which is also called the cost function or, with reverse sign, the fitness function)

$$\varphi(a(c)) = \sum_c f(c) \sum_{c'} p(c'|c) d(a(c), a(c')) \quad (1)$$

Here codon $c$ codes for amino acid $a$ or stop-codon (such mapping is dependent on a code), $f(c)$ is the frequency of codon $c$, $p(c'|c)$ corresponds to the conditional probability for codons $c$ and $c'$, and $d(a(c), a(c'))$ is the cost associated with exchange of amino acids $a(c)$ and $a(c')$. The minimum of stability function (1) corresponds to the most robust code. The choice of functions $f(c)$, $p(c'|c)$, and $d(a(c), a(c'))$ may affect the details of stability assessment (Haig and Hurst, 1991; Freeland and Hurst, 1998; Goodarzi et al., 2005; Novozhilov et al., 2007). In particular, the inclusion of codon frequency $f(c)$ deteriorates commonly the stability (Zhu et al., 2003; Archetti, 2004). On the contrary, the proper choice of $p(c'|c)$ and $d(a(c), a(c'))$ may significantly improve the estimates.

For our purposes we chose the unbiased variant proposed by Haig and Hurst (1991), $f(c) = $ const, $p(c'|c) = 1/9$ if codons $c$ and $c'$ differ by one-nucleotide replacement and $p(c'|c) = 0$ otherwise, and

$$d(a(c), a(c')) = (P(a(c)) - P(a(c')))^2 \quad (2)$$

where $P(a(c))$ is a physicochemical characteristic of amino acid. We used the hydrophobicity $H$ (Black and Mould, 1991) and the molecular volume $V$ (Zamjatnin, 1972) as particular characteristics, because these factors play the most important role in the folding of proteins. The hydrophobicity and molecular volume scales were mapped onto the interval (0, 1). The joint influence of these properties was assessed with the sum of the corresponding stability functions

$$\varphi_S \equiv \varphi_H + \varphi_V \qquad (3)$$

The stability function $\varphi_S$ is similar to the sum over squared distances between amino acids introduced by Miyata et al. (1979). The relevance of this distance to the genetic code was noted by Di Giulio (1989) and Cavalcanti et al. (2000).

A more complicated conditional probability simulating the actual mechanisms of translation errors (Parker, 1989; Ogle and Ramakrishnan, 2005) as well as the bias between transition (C ↔ U and A ↔ G) and transversion (C, U ↔ A, G) mutations was proposed by Freeland and Hurst (1998). In this variant the conditional probability is defined as

$$p(c'|c) = \begin{cases} 1/N & \text{if } c' \text{ and } c \text{ differ in the 3rd base only} \\ 1/N & \text{if } c' \text{ and } c \text{ differ in the 1st base only} \\ & \text{and cause a transition} \\ 0.5/N & \text{if } c' \text{ and } c \text{ differ in the 1st base only} \\ & \text{and cause a transversion} \\ 0.5/N & \text{if } c' \text{ and } c \text{ differ in the 2nd base only} \\ & \text{and cause a transition} \\ 0.1/N & \text{if } c' \text{ and } c \text{ differ in the 2nd base only} \\ & \text{and cause a transversion} \end{cases}$$

Factor $N$ ensures the proper normalization of conditional probability $\sum_{c'} p(c'|c) = 1$ for any codon $c$ and is equal to 5.7 in this model. We used also this variant for the comparison purposes. The unbiased stability function characterizes the general stability of the genetic code, whereas the biased stability function reflects more properly the underlying molecular mechanisms.

It seems to be natural to prescribe the zero values of physicochemical parameters to the stop-codons. The other possibility consists in prescribing to stop-codons the mean value over close sense codons related to stop-codons (Goodarzi et al., 2004). To avoid a confusion of such definition with mapping physicochemical parameters onto the interval (0, 1), these parameters were remapped onto the interval (0.1, 1) in the scheme with stop-codons. We have checked that remapping onto the narrower interval (0.2, 1) weakly affected the results. Unlike the scheme with stop-codons, the results in the scheme without stop-codons do not depend on the linear remapping of a scale.

The relative robustness of a code obtained under delin or swap modifications of the standard code was assessed via difference

$$\Delta \varphi = \varphi_{\text{modified}} - \varphi_{\text{standard}} \qquad (4)$$

Negative variations $\Delta\varphi$ correspond to more robust codes in comparison with the standard one. Below we will always use the normalized variations of stability function,

$$\Delta \tilde{\varphi} = \Delta \varphi / \sigma(\Delta \varphi) \qquad (5)$$

where $\sigma(\Delta\varphi)$ is the standard deviation calculated for all variations produced by modifications of the standard code.

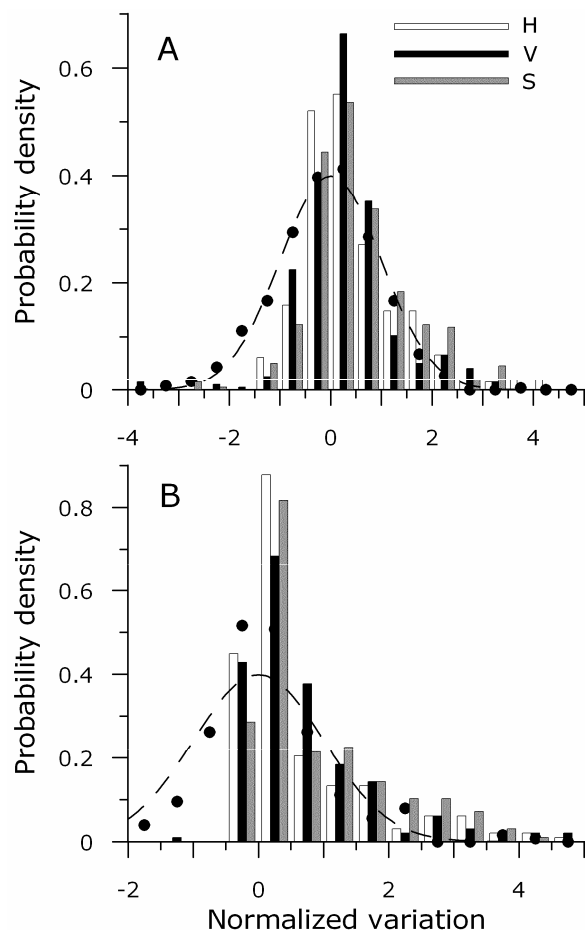

Fig. 1. The distribution of normalized variations of stability function produced by delins (A) and swaps (B) in the standard code. The stop-codons are discarded. The broken line shows the Gaussian distribution. The circles correspond to the distribution of normalized variations for a particular random code with the same number of codons in the sets coding for different amino acids as in the standard code. In this and other figures the distributions correspond to the unbiased conditional probability (Haig and Hurst, 1991). In the scheme with biased conditional probability (Freeland and Hurst, 1998) the drop in negative variations is even sharper.

### 2.3. Distribution of variations of stability function

The distributions of normalized variations of stability function under delin and swap modifications of the standard code are shown in Figs. 1 and 2, whereas the quantitative characteristics of negative variations related to the more robust codes are summarized in Table 1. The fraction of variations of interest is defined as

$$f_{\text{observed}} = N_{\text{observed}} / N_{\text{total}} \qquad (6)$$

where $N_{\text{total}}$ is the total number of modifications and $N_{\text{observed}}$ is the number of modifications with variations of stability function less than a given threshold value. The suppression of negative variations beyond the range of one standard deviation was proved with the stricter criterion for the relative fraction of variations exceeding this threshold (i.e., with the ratio of variations with sweeps exceeding one standard deviation to the total number of negative variations).

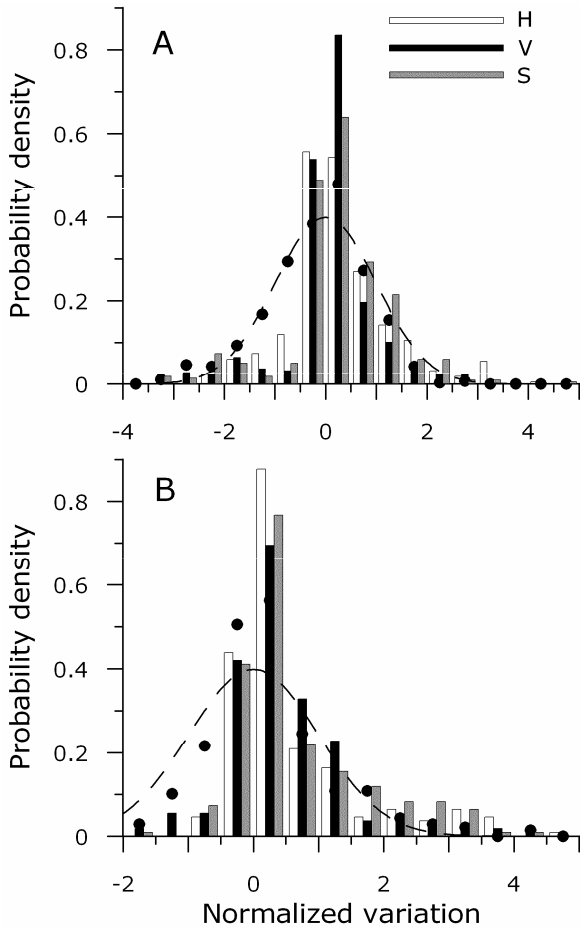

Fig. 2. The distribution of normalized variations of stability function produced by delins (A) and swaps (B) in the standard code with stop-codons. The broken line shows the Gaussian distribution. The circles correspond to the distribution of normalized variations for a particular random code with the same number of codons in the sets coding for different amino acids or stop-codons as in the standard code.

The corresponding mean fractions for the random codes with the same number of codons in the sets coding for different amino acids or stop-codons are 0.5 and 0.32, respectively. The statistical significance of observable deviations from these values may be assessed with $z$-value (Johnson and Leone, 1977; Weir, 1990)

$$z_f = (\bar{f}_{random} - f_{observed})/\sigma(\bar{f}_{random}) \qquad (7)$$

$$\sigma^2(\bar{f}_{random}) = \bar{f}_{random}(1 - \bar{f}_{random})/N_{total} \qquad (8)$$

where $\bar{f}_{random}$ is the mean frequency for the random codes. In our conditions the statistics of $z$-values may be approximated by the Gaussian distribution. Except few examples, the deviations of observed frequencies in Table 1 from the expected values for random codes correspond to $z$-values in the range 3.5–8.0, which are highly statistically significant (cf., $z = 2.58$ for $Pr = 0.01$). This gives evidence of the distinctly non-random nature of the standard code in accordance with the previous observations (Haig and Hurst, 1991; Freeland and Hurst, 1998; Goodarzi et al., 2005; Novozhilov et al., 2007). The comparison of Figs. 1 and 2 as well as data in Table 1 shows the invariably lower robustness of the code with stop-codons (see also Section 5 below).

The applicability of Gaussian criteria for random codes was checked by control simulations with the several runs of $10^3$ patterns. The statistics of variations of the stability function produced by delin operations matches Gaussian distribution in average both for the whole ensemble of random codes and for the particular representatives (see panels A in Figs. 1 and 2). The random variations produced by swaps may deviate a little from Gaussian statistics (or reveal the slow convergence). The mean relative fraction of negative variations exceeding one standard deviation turned out to be 0.27 ± 0.01 for swaps in the random codes (in comparison with 0.32 for Gaussian distribution). The recalculation of $z$-values with this frequency does not change the resulting conclusion on the high significance of observed deviations for the standard code from the random patterns.

The amino acids encoded by the genetic code possess several physicochemical characteristics simultaneously (Haig and Hurst, 1991; Xia and Li, 1998). Interestingly, the application of joint stability function (3) often improves the robustness. This feature is inherent to the standard code and is absent in random codes.

The more detailed information on the replacements in the second codon position for modifications of the standard code with negative variations of stability function in the scheme with excluded stop-codons is presented in Tables 2 and 3. The second codon position was chosen as most informative (see also below). The corresponding matrices were obtained as follows. For example, the choice of stability function $\varphi_H$ and deletion of the codon AGA from the set coding for arginine and insertion it to the set coding for glycine with close codon GGA leads to the more robust code (or negative variation (4)). Such delin operation is denoted AGA $\rightarrow$ GGA and gives +1 contribution to the corresponding matrix element G $\rightarrow$ G in Table 2. For the same function $\varphi_H$ the swap of codons AAA $\leftrightarrow$ AAC also produces negative variation of stability function and gives +1 contribution to the corresponding matrix element A $\leftrightarrow$ A. Unlike symmetric matrix for swaps, the matrix for delins is non-symmetric, i.e. all non-diagonal elements are significant.

**3. Stability of binary blocks of the standard code**

It is well known that binary blocks in the standard code (i.e., NRN-NYN, NWN-NSN, and NMN-NKN) code for amino acids with different physicochemical properties (Rumer, 1968; Jungck, 1978; Wolfenden et al., 1979; Blalock and Smith, 1984; Taylor and Coates, 1989; Chechetkin, 2003; Wilhelm and Nikolaeva, 2004). Here R = (A, G) and Y = (C, U), W = (A, U) and S = (C, G), K = (G, U) and M = (A, C), and N is any nucleotide. The block NRN codes for the amino acids with lower hydrophobicity,

**Table 1**

Characteristics of negative normalized variations of stability function for modifications of the standard code

A. The unbiased conditional probability

| Operation | Total | Fraction with Δφ < 0 | | | Relative fraction with Δφ < −1 | | | Stop-codons |
|---|---|---|---|---|---|---|---|---|
| | | $\varphi_H$ | $\varphi_V$ | $\varphi_S$ | $\varphi_H$ | $\varphi_V$ | $\varphi_S$ | |
| Delin | 438 | 0.42 | 0.38 | 0.36 | 0.19 | 0.25 | 0.24 | Yes |
| Delin | 392 | 0.37 | 0.35 | 0.32 | 0.08 | 0.10 | 0.11 | No |
| Swap | 219 | 0.24 | 0.29 | 0.25 | 0.00 | 0.19 | 0.02 | Yes |
| Swap | 196 | 0.22 | 0.22 | 0.14 | 0.00 | 0.02 | 0.00 | No |

B. The biased conditional probability

| Operation | Total | Fraction with Δφ < 0 | | | Relative fraction with Δφ < −1 | | | Stop-codons |
|---|---|---|---|---|---|---|---|---|
| | | $\varphi_H$ | $\varphi_V$ | $\varphi_S$ | $\varphi_H$ | $\varphi_V$ | $\varphi_S$ | |
| Delin | 438 | 0.32 | 0.34 | 0.32 | 0.21 | 0.27 | 0.30 | Yes |
| Delin | 392 | 0.28 | 0.31 | 0.27 | 0.00 | 0.07 | 0.06 | No |
| Swap | 219 | 0.20 | 0.25 | 0.19 | 0.00 | 0.19 | 0.02 | Yes |
| Swap | 196 | 0.16 | 0.12 | 0.07 | 0.00 | 0.00 | 0.00 | No |

For random codes, the expected fractions for Δφ < 0 and Δφ < −1 are 0.5 and 0.32, respectively.
Relative fraction with Δφ < −1 is defined with respect to the number of variations with Δφ < 0.

whereas the block NWN codes for the amino acids with larger side-chain volumes. The coding of amino acids with opposite properties, which is most robust to translation errors and point mutations, needs the fixation of the counterpart nucleotides in the same positions within codons for the each binary block coding for amino acids with approximately homogeneous properties (Chechetkin, 2003). For example, in the case of binary code R-Y the corresponding optimal subdivisions should be of the form, RNN-YNN, or NRN-NYN, or NNR-NNY. The fixation of the most informative nucleotide in the middle of the codons ensures the most reliable molecular recognition during translation. This mode of coding also provides the approximate stability of hydrophobic or hydrophilic stretches with respect to frameshifts (Chechetkin, 2003). The hydrophobicity of purines R is higher than that of pyrimidines Y and anticorrelates with the lower/higher hydrophobicity of encoded amino acids. The selection of amino acids with the opposite physicochemical characteristics allows proper folding of globular and transmembrane proteins and ensures their structural variety. Amino acids with such approximately dichotomous physicochemical properties are actually encoded in the genetic code.

We studied the robustness of coding within separate binary blocks of the standard code. This means, e.g., that NRN and NYN are considered as the separate codes coding for about 10 amino acids or stop-codons. We used the unbiased stability function for such an analysis to alleviate the comparison with the whole code. The resulting distributions of the normalized variations of unbiased stability function in the scheme of modifications without stop-codons are presented in Figs. 3 and 4. The details of related distributions are summarized in Table 4. Remarkably, the suppression of negative variations and their sweeps beyond one standard deviation is clearly seen in the separate binary blocks of the standard code as well. The statistical significance of the observed deviations from the expected frequencies in the counterpart random codes can be proved with $z$-criterion (see Eq. (7)). The corresponding fractions of variations with Δφ < 0 and Δφ < −1 turn out to be comparable for the whole code and for its binary blocks (cf. Tables 1A and 4). The inclusion of stop-codons deteriorates the robustness of coding in the corresponding blocks similarly to the whole code (cf. Fig. 2).

The distribution of matrix elements in the binary matrices of replacements for the whole code may serve as an indicator of the relationship between binary block structure and the higher stability of the genetic code. The relevant binary matrices are shown in Table 5. If delins or swaps transpose the codon(s) from the sets belonging to the different binary blocks, then such modification violates the optimal binary subdivision and should be expected to deteriorate the robustness of the code. This means the negative bias for the non-diagonal elements in the binary matrices of replacements corresponding to the more robust codes (Table 5). As the first example, we consider the choice of stability function $\varphi_H$. The blocks NRN and NYN code for hydrophilic and hydrophobic amino acids, respectively. Such subdivision of amino acids holds for NRN-NYN blocks and fails for the other binary blocks, NSN-NWN and NKN-NMN. As is seen from Table 5 (line *H*), the ratio of the sum for non-diagonal elements to the sum of all matrix elements is the lowest for R-Y matrix in comparison with S-W and K-M matrices for delins. Because of the strict symmetry the relevant ratio in matrices for swaps must include only one of non-diagonal elements. The corresponding mean ratios for the random codes would be equal to $2/9 \approx 0.22$. The observed values for R-Y matrices are 0.17 (delins) and 0.07 (swaps). As the second example, let us consider the choice of stability function $\varphi_V$. The subdivision of amino acids with different molecular volumes is most significant in NSN-NWN blocks. Therefore, the lowest ratio for non-diagonal elements should be obtained for S-W matrix in accordance with data in Table 5 (line $V$).

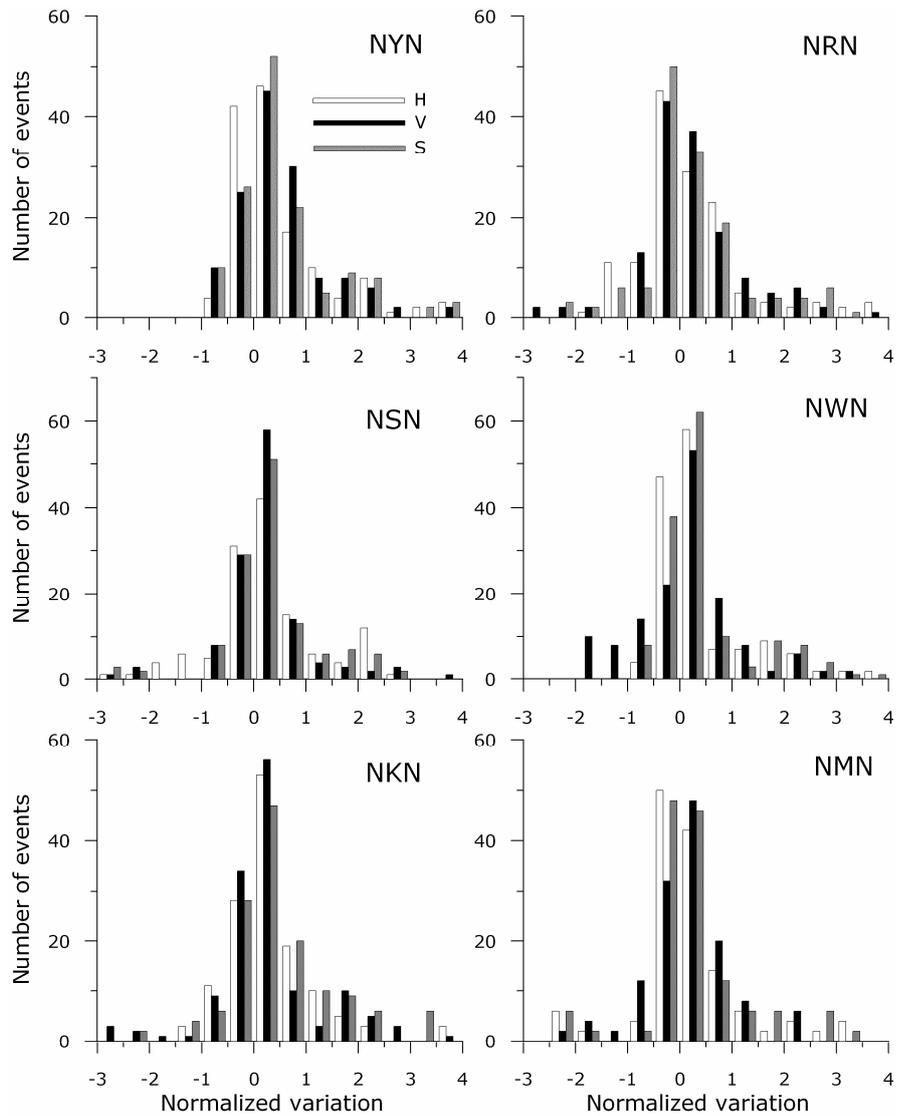

Fig. 3. The distribution of normalized variations of stability function produced by delins in the different binary blocks of the standard code. The stop-codons are discarded. Each binary block is considered to be the separate code coding for about ten amino acids. Here R = (A, G) and Y = (C, U), W = (A, U) and S = (C, G), K = (G, U) and M = (A, C), and N is any nucleotide.

The other important binary subdivision of the genetic code is related to class I and class II aminoacyl-tRNA synthetases (AARS) (Woese et al., 2000; Ribas de Pouplana and Schimmel, 2001; Delarue, 2007; Rodin and Rodin, 2006, 2008). The codons NUN and NCN code for amino acids associated with class I and class II AARS (except two codons UUU and UUC coding for phenylalanine), while the codons NGN and NAN may be associated with the two classes with discernibly lower significance, resulting in approximate subdivision NKN-NMN. The error minimization theory proved to be efficient for subdivision by the two classes of AARS as well (Cavalcanti et al., 2000; Chechetkin, 2006). The corresponding data on the local stability of binary blocks related to AARS are also presented in Table 4.

The binary blocks may be approximately considered as subcodes within the standard code. Their robustness supports this suggestion. The further binary subdivisions within binary blocks are also plausible.

## 4. Stability of the standard code and amino acids

The transposition of a codon from one coding set to the other via delin operation, $c \to c'$, may be treated as the replacement of corresponding encoded amino acids, $a(c) \to a'(c')$. If an operation produces a more robust code in comparison with the standard one, the natural weighting for the assessment of potential preference

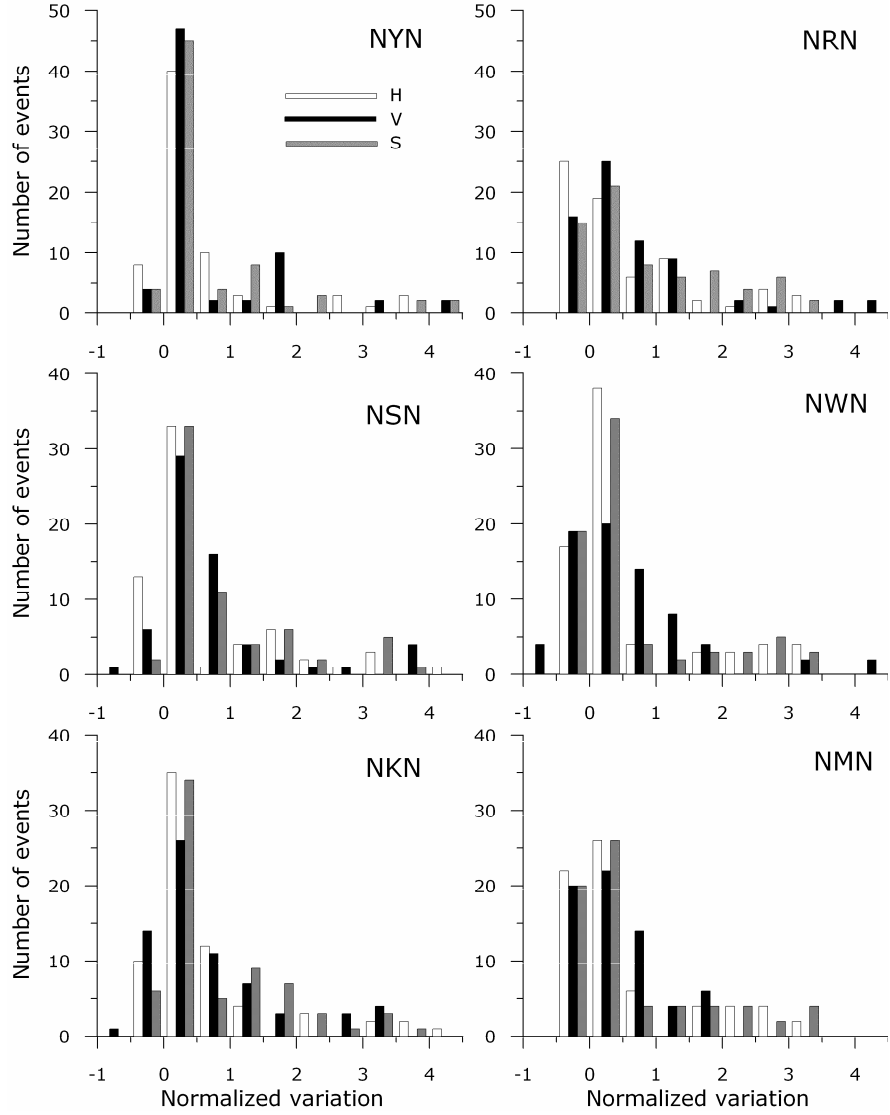

Fig. 4. The distribution of normalized variations of stability function produced by swaps in the different binary blocks of the standard code. The stop-codons are discarded. Each binary block is considered to be the separate code coding for about ten amino acids. Here R = (A, G) and Y = (C, U), W = (A, U) and S = (C, G), K = (G, U) and M = (A, C), and N is any nucleotide.

caused by such modification of the code may be defined as

$$w_{aa'}(c \to c') = \left(\frac{2}{\pi}\right)^{1/2} \int_0^{|\Delta\varphi|} dx\, e^{-x^2/2} \quad (9)$$

where $|\Delta\varphi|$ is the modulus of respective normalized variation (5). The summation over all delin operations $c \to c'$ producing more robust codes and associated with given replacement $a \to a'$ generates the matrix

$$T_{aa'} = \sum_{c,c'} w_{aa'}(c \to c') \quad (10)$$

An example of amino acid replacement matrix in the scheme with the stop-codons and the choice of unbiased stability function $\varphi_S$ is given in Table 6. For convenience the resulting matrix elements were mapped onto the interval (0, 1). The matrix $\hat{T}$ could be used for simulation of molecular evolution, which would be produced by the more robust variations of the standard code. The relevant scenario may be formulated in terms of Markovian matrix

$$\hat{M} = \hat{D} + r\hat{T} \quad (11)$$

where $\hat{D}$ is the diagonal matrix and $r$ characterizes the small rate of replacements and should be introduced separately. Markovian matrix needs the fulfillment of conditions

$$D_{aa} + r\sum_{a'} T_{aa'} = 1 \quad (12)$$

determining the diagonal matrix elements $D_{aa}$ at given $r$ and $\hat{T}$. Formally, the matrix $\hat{M}$ is similar to 1 PAM

**Table 2**
The number of replacements in the second codon position for modifications of the standard code with negative variations of stability function (the stop-codons are excluded)

A. The unbiased conditional probability

Delins

|   | A | G | U | C |   | A | G | U | C |   | A | G | U | C |
|---|---|---|---|---|---|---|---|---|---|---|---|---|---|---|
| A | 28 | 8 | 0 | 6 | A | 34 | 2 | 8 | 2 | A | 34 | 2 | 0 | 6 |
| G | 6 | 19 | 1 | 6 | G | 12 | 19 | 11 | 11 | G | 8 | 16 | 2 | 11 |
| U | 2 | 3 | 25 | 7 | U | 2 | 0 | 16 | 0 | U | 2 | 0 | 22 | 2 |
| C | 0 | 6 | 0 | 28 | C | 2 | 2 | 0 | 16 | C | 0 | 2 | 0 | 18 |
|   |   | H |   |   |   |   | V |   |   |   |   | S |   |   |

Swaps

|   | A | G | U | C |   | A | G | U | C |   | A | G | U | C |
|---|---|---|---|---|---|---|---|---|---|---|---|---|---|---|
| A | 15 | 7 | 0 | 2 | A | 10 | 5 | 4 | 0 | A | 10 | 5 | 0 | 0 |
| G | 7 | 4 | 0 | 1 | G | 5 | 1 | 6 | 4 | G | 5 | 0 | 1 | 1 |
| U | 0 | 0 | 5 | 0 | U | 4 | 6 | 9 | 0 | U | 0 | 1 | 4 | 0 |
| C | 2 | 1 | 0 | 10 | C | 0 | 4 | 0 | 4 | C | 0 | 1 | 0 | 7 |
|   |   | H |   |   |   |   | V |   |   |   |   | S |   |   |

B. The biased conditional probability

Delins

|   | A | G | U | C |   | A | G | U | C |   | A | G | U | C |
|---|---|---|---|---|---|---|---|---|---|---|---|---|---|---|
| A | 38 | 2 | 0 | 2 | A | 26 | 2 | 8 | 2 | A | 38 | 2 | 0 | 2 |
| G | 7 | 17 | 0 | 3 | G | 11 | 17 | 11 | 11 | G | 7 | 13 | 1 | 7 |
| U | 2 | 1 | 25 | 0 | U | 0 | 0 | 14 | 0 | U | 2 | 0 | 22 | 0 |
| C | 0 | 2 | 0 | 12 | C | 2 | 2 | 0 | 15 | C | 0 | 2 | 0 | 8 |
|   |   | H |   |   |   |   | V |   |   |   |   | S |   |   |

Swaps

|   | A | G | U | C |   | A | G | U | C |   | A | G | U | C |
|---|---|---|---|---|---|---|---|---|---|---|---|---|---|---|
| A | 14 | 7 | 0 | 2 | A | 2 | 4 | 2 | 0 | A | 6 | 2 | 0 | 0 |
| G | 7 | 4 | 0 | 1 | G | 4 | 1 | 6 | 3 | G | 2 | 0 | 1 | 0 |
| U | 0 | 0 | 3 | 0 | U | 2 | 6 | 6 | 0 | U | 0 | 1 | 4 | 0 |
| C | 2 | 1 | 0 | 0 | C | 0 | 3 | 0 | 0 | C | 0 | 0 | 0 | 0 |
|   |   | H |   |   |   |   | V |   |   |   |   | S |   |   |

**Table 3**
The number of replacements in the second codon position for modifications of the standard code with negative variations of cost function beyond one standard deviation (the stop-codons are excluded)

A. The unbiased conditional probability

Delins

|   | A | G | U | C |   | A | G | U | C |   | A | G | U | C |
|---|---|---|---|---|---|---|---|---|---|---|---|---|---|---|
| A | 0 | 0 | 0 | 0 | A | 4 | 0 | 0 | 0 | A | 2 | 2 | 0 | 0 |
| G | 2 | 8 | 0 | 2 | G | 1 | 5 | 2 | 2 | G | 1 | 5 | 1 | 3 |
| U | 0 | 0 | 0 | 0 | U | 0 | 0 | 0 | 0 | U | 0 | 0 | 0 | 0 |
| C | 0 | 0 | 0 | 0 | C | 0 | 0 | 0 | 0 | C | 0 | 0 | 0 | 0 |
|   |   | H |   |   |   |   | V |   |   |   |   | S |   |   |

The only corresponding replacement for swaps is the unique replacement G ↔ U for V-matrix.

B. The biased conditional probability

Delins

|   | A | G | U | C |   | A | G | U | C |   | A | G | U | C |
|---|---|---|---|---|---|---|---|---|---|---|---|---|---|---|
| A | 0 | 0 | 0 | 0 | A | 0 | 0 | 0 | 0 | A | 0 | 0 | 0 | 0 |
| G | 0 | 0 | 0 | 0 | G | 0 | 5 | 1 | 3 | G | 0 | 5 | 0 | 1 |
| U | 0 | 0 | 0 | 0 | U | 0 | 0 | 0 | 0 | U | 0 | 0 | 0 | 0 |
| C | 0 | 0 | 0 | 0 | C | 0 | 0 | 0 | 0 | C | 0 | 0 | 0 | 0 |
|   |   | H |   |   |   |   | V |   |   |   |   | S |   |   |

The corresponding replacements for swaps are absent.

**Table 4**

Characteristics of negative normalized variations of stability function for modifications of binary blocks in the standard code in the scheme with unbiased conditional probability and excluded stop-codons

| Block | Operation | Total | Fraction with Δφ < 0 | | | Relative fraction with Δφ < −1 | | |
|---|---|---|---|---|---|---|---|---|
| | | | $\varphi_H$ | $\varphi_V$ | $\varphi_S$ | $\varphi_H$ | $\varphi_V$ | $\varphi_S$ |
| NRN | Delin | 138 | 0.49 | 0.46 | 0.49 | 0.18 | 0.10 | 0.16 |
| | Swap | 69 | 0.36 | 0.23 | 0.22 | 0.00 | 0.00 | 0.00 |
| NYN | Delin | 138 | 0.33 | 0.62 | 0.26 | 0.00 | 0.00 | 0.00 |
| | Swap | 69 | 0.12 | 0.06 | 0.06 | 0.00 | 0.00 | 0.00 |
| NSN | Delin | 128 | 0.38 | 0.34 | 0.34 | 0.25 | 0.14 | 0.14 |
| | Swap | 64 | 0.20 | 0.11 | 0.03 | 0.00 | 0.00 | 0.00 |
| NWN | Delin | 146 | 0.35 | 0.37 | 0.32 | 0.00 | 0.33 | 0.00 |
| | Swap | 73 | 0.23 | 0.32 | 0.26 | 0.00 | 0.00 | 0.00 |
| NKN | Delin | 138 | 0.30 | 0.36 | 0.29 | 0.07 | 0.14 | 0.15 |
| | Swap | 69 | 0.14 | 0.22 | 0.09 | 0.00 | 0.00 | 0.00 |
| NMN | Delin | 136 | 0.46 | 0.38 | 0.43 | 0.13 | 0.15 | 0.14 |
| | Swap | 68 | 0.32 | 0.29 | 0.29 | 0.00 | 0.00 | 0.00 |
| Class I | Delin | 84 | 0.24 | 0.26 | 0.24 | 0.00 | 0.11 | 0.02 |
| | Swap | 42 | 0.21 | 0.29 | 0.10 | 0.00 | 0.00 | 0.12 |
| Class II | Delin | 112 | 0.30 | 0.25 | 0.25 | 0.02 | 0.00 | 0.00 |
| | Swap | 56 | 0.11 | 0.07 | 0.07 | 0.00 | 0.00 | 0.00 |

For random codes, the expected fractions for Δφ < 0 and Δφ < −1 are 0.5 and 0.32, respectively. Relative fraction with Δφ < −1 is defined with respect to the number of variations with Δφ < 0.

(Henikoff and Henikoff, 1992; Benner et al., 1994). Powering $\hat{M}$ simulates the underlying molecular evolution.

As the numbers of direct and reverse operations $a \rightarrow a'$ and $a' \rightarrow a$ leading to the more robust codes are not equal to each other (the same refers to the corresponding sweeps in variations of stability function), the matrix $\hat{T}$ appears to be non-symmetric, $T_{aa'} \neq T_{a'a}$. The integral parameter determined for a given amino acid $a$,

$$\Delta T_a = \sum_{a'} \left( T_{a'a} - T_{aa'} \right) \quad (13)$$

characterizes approximately the difference between "gain" and "loss" produced by all delin operations that are related to this amino acid and lead to more robust modifications of the standard code. In the scheme with stop-codons the contribution of operations $a \rightarrow Ter$ and $Ter \rightarrow a$ should also be taken into account.

Ranking amino acids according to the parameter $\Delta T_a$ defined by Eq. (13) and the data in Table 6 (with addition of contributions $a \rightarrow Ter$ and $Ter \rightarrow a$) yields the order, C > N > S > H > T > V > Q > M > A > P > D > L > G > K > E > I > F > R > W > Y, where the underlined amino acids correspond to the negative values of $\Delta T_a$.

At the present stage of molecular evolution there exists a strong bias against variations in the genetic code (Crick, 1968; Freeland et al., 2000). If the selection of more robust codes would still be one of the driving forces of molecular evolution, then the ranked order obtained above might serve for the assessment of the actual rates of gain/loss of particular amino acids. Jordan et al. (2005) suggested the existence of the universal trend in the loss of amino acids which were first incorporated into the genetic code and the gain of amino acids which were probably be recruited later (see also the discussion by Hurst et al., 2006). Ranking amino acids according to the estimated rates of gain/loss (Table 3 in the paper by Jordan et al., 2005) would produce the order, S > V > T > I > M > H > N > C > F > R > Q > W > Y > L > D > G > K > E > P > A, the underlined amino acids correspond to the negative (loss) rates. Spearman correlations between two ranked sequences turned out to be 0.44 (Pr = 0.05). The correlation of this kind remains the intrigue whether at least part of amino acid replacements may be due to the more stable mode of coding or not.

At the end of this section we discuss how the different factors affect the rank of amino acid according to the parameter $\Delta T_a$. (i) The choice of $\varphi_H$ or $\varphi_V$ yields the independent sequences (rank correlation coefficient $R \approx 0.2$), yet both sequences reveal significant correlations with the pattern obtained with $\varphi_S$ ($R \approx 0.7$–0.8). (ii) The ranks are not very sensitive to the definition of weights in Eq. (9). The comparison of results for the choice $w_{aa'} = 1$, |Δφ| (linear), and integral weights defined by Eq. (9) showed the correlations $R \approx 0.7$ between $w_{aa'} = 1$ and linear/integral weights, and $R \approx 0.9$ between linear and

**Table 5**
Representation of replacements in Table 2A in terms of binary codes

Delins

|   | R  | Y  |   | W  | S  |   | K  | M  |
|---|----|----|---|----|----|---|----|----|
| R | 61 | 13 | W | 55 | 24 | K | 48 | 21 |
| Y | 11 | 60 | S | 7  | 59 | M | 14 | 62 |

H

|   | R  | Y  |   | W  | S  |   | K  | M  |
|---|----|----|---|----|----|---|----|----|
| R | 67 | 32 | W | 60 | 4  | K | 46 | 25 |
| Y | 6  | 32 | S | 25 | 48 | M | 12 | 54 |

V

Swaps

|   | R  | Y  |   | W  | S  |   | K | M  |
|---|----|----|---|----|----|---|---|----|
| R | 26 | 3  | W | 20 | 9  | K | 9 | 8  |
| Y | 3  | 15 | S | 9  | 15 | M | 8 | 27 |

H

|   | R  | Y  |   | W  | S  |   | K  | M  |
|---|----|----|---|----|----|---|----|----|
| R | 16 | 14 | W | 23 | 11 | K | 16 | 13 |
| Y | 14 | 13 | S | 11 | 9  | M | 13 | 14 |

V

integral weighting. (iii) The schemes with and without stop-codons reveal close ranking ($R \approx 0.9$). (iv) The schemes with unbiased and biased conditional probabilities also provide close ranking ($R \approx 0.8$–$0.9$). Nevertheless, powering counterpart matrices should produce more divergent results in view of sensitivity of this operation on the choice of initial matrix. The choice of unbiased stability function $\varphi_S$, inclusion of stop-codons, and integral weighting defined by Eq. (9) yielded the highest correlations with the rates obtained by Jordan et al. (2005).

The replacement matrix $\hat{T}$ corresponding to swaps is symmetric and does not contribute into $\Delta T_a$. The potential application of these results to the analysis of protein sequences deserves separate investigation.

### 5. Stability and variants of the genetic code

Despite the strong bias against variations in the standard code, there are known more than twenty alternatives of the genetic code (Knight et al., 2001; Santos et al., 2004). Taking the data provided by Knight et al. (2001), we compared the known modes of codon reassignment with the corresponding delin and swap modifications of the standard code (see Table 7). All results in Table 7 correspond to the scheme with stop-codons and biased stability function. Three cases of codon reassignment referring to two-point difference in the codons were considered separately. The data for the set CUN correspond to the reassignment of all codons to threonine.

The stop-codons deteriorate invariably the stability of the genetic code and are among the most frequent reassigned codons. Only in the vertebrate and *Thraustochytrium* mitochondrial codes the number of stop-codons increases to four from three in the standard code. Commonly, their number decreases to two or even one in the majority of known variants. The molecular mechanisms concerning the reassignment of stop-codons were discussed by Knight et al. (2001), Ivanov et al. (2001), and Santos et al. (2004).

Unlike stop-codons, nearly all delin operations with the initiation codon AUG → $c'$ do not yield the more robust codes (cf. Table 6). This feature may be considered as a bias ensuring the more reliable start of translation. Similar bias works also on the level of selection of tRNA anticodons (Chechetkin, 2006).

The results in Table 7 indicate that the modifications leading to the more robust codes may be preferential during selection and may provoke the development of alternative codes. The detailed comparative analysis of the available code variants may shed additional light upon the origin and evolution of the genetic code.

### 6. Discussion

The search of the most optimal code associated with the global minimum of stability function (1) is incomplete without formulation of additional constraints. Stability function serves as an indicator of the minimization efficiency of a code with respect to translation errors and point mutations and characterizes the homogeneity of coding for the physicochemical parameters. Without any constraints the global minimum of stability function would be attained by collecting all codons in one coding set.

The other important aspect related to the origin of the genetic code (and missed in stability function) is the problem of diversity or complexity (reviewed by Abel and Trevors, 2006). Self-reproducing needs some critical level of complexity. For example, the coding might start from a unique most ancient amino acid and corresponding codon(s). Though conceivable from the general positions, such a hypothesis does not satisfy the complexity

**Table 6**

The matrix of amino acids replacements induced by delin operations in the standard code, which lead to more robust coding.
The data correspond to the scheme with stop-codons and the choice of unbiased stability function $\varphi_S$

|     | A    | C    | D    | E    | F | G    | H    | I    | K    | L    | M    | N    | P    | Q    | R    | S    | T    | V    | W    | Y    | Ter  |
|-----|------|------|------|------|---|------|------|------|------|------|------|------|------|------|------|------|------|------|------|------|------|
| A   |      |      |      |      |   |      |      |      |      |      |      |      |      |      |      |      | 0.01 |      |      |      |      |
| C   |      |      |      |      |   |      |      |      |      |      |      |      |      |      |      |      |      |      |      |      |      |
| D   | 0.03 |      |      | 0.07 |   |      | 0.14 |      |      |      |      | 0.24 |      |      |      |      |      |      |      |      |      |
| E   |      |      | 0.09 |      |   |      |      |      |      |      |      |      |      | 0.10 |      |      |      |      |      |      |      |
| F   |      |      |      |      |   |      |      | 0.13 |      | 0.38 |      |      |      |      |      |      |      | 0.15 |      | 0.04 |      |
| G   | 0.14 |      |      |      |   |      |      |      |      |      |      |      |      |      | 0.11 |      |      |      |      |      |      |
| H   |      |      |      |      |   |      |      |      |      |      |      |      |      | 0.15 |      |      |      |      |      |      |      |
| I   |      |      |      |      |   |      |      |      | 0.21 |      |      |      |      |      |      |      |      | 0.17 |      |      |      |
| K   |      |      |      | 0.02 |   |      |      |      |      |      |      | 0.20 |      | 0.10 |      | 0.01 |      |      |      |      |      |
| L   |      |      |      |      |   |      |      |      |      |      | 0.13 |      | 0.04 |      |      |      |      | 0.29 |      |      |      |
| M   |      |      |      |      |   |      |      |      |      |      |      |      |      |      |      |      |      |      |      |      |      |
| N   |      |      |      |      |   |      |      | 0.05 |      |      |      |      |      |      |      | 0.05 |      |      |      |      |      |
| P   |      |      |      |      |   |      |      |      |      |      |      |      |      |      |      | 0.12 |      |      |      |      |      |
| Q   |      |      |      |      |   |      |      |      |      |      |      |      |      |      |      |      |      |      |      |      |      |
| R   |      | 0.16 |      |      |   | 0.04 | 0.30 |      | 0.30 |      | 0.07 |      |      | 0.16 | 0.26 |      | 0.83 | 0.37 |      |      |      |
| S   | 0.14 | 0.19 |      |      |   |      |      |      |      |      |      | 0.10 | 0.19 |      |      |      | 0.32 |      |      |      |      |
| T   |      |      |      |      |   |      |      |      |      |      |      |      |      |      |      |      |      |      |      |      |      |
| V   |      |      |      |      |   |      |      |      |      |      |      |      |      |      |      |      |      |      |      |      |      |
| W   |      | 0.68 |      |      |   | 0.34 |      |      | 0.28 |      |      |      |      |      | 0.67 | 0.34 |      |      |      |      | 0.66 |
| Y   |      | 0.62 | 0.62 |      |   |      | 0.63 |      |      |      | 0.66 |      |      |      |      | 0.66 |      |      |      |      | 0.40 |
| Ter |      | 0.66 |      | 0.65 |   | 0.31 |      | 0.67 | 0.76 |      |      |      |      | 0.67 | 0.55 | 1.00 |      |      | 0.34 | 0.97 |      |

restriction. The coexistence of expanded sets coding for leucine, serine and arginine with unique codons coding for tryptophan and methionine may be treated as the compromise between coding stability and diversity. The constraints imposed by the binary block structure of the genetic code needs further detailing for the proper formulation of the error-minimization problem (see also Chechetkin, 2003). The formulation of genetically meaningful constraints related to the origin and evolution of the genetic code, which characterize the critical diversity or complexity, remains still the unsolved problem.

Commonly, the problem of constraints is circumvented by fixing numbers of codons in the coding sets in the same mode as in the standard code. In some cases this condition is too restrictive and does not agree with the available variants of the genetic code. In our approach, we assumed that the constraints are not important in the close vicinity of the standard code. This assumption cannot be ignored at the subsequent iteration steps for the search of global minimum of stability function. Such approach allows investigating impact of the small modifications on the stability of the genetic code.

The results appeared to be informative and possess the predictive power. The joint action of random factors and regular selection is clearly seen in the character of stability of the standard code. The randomness is expressed in the unexpectedly high number of modifications leading to the more robust codes. Their number and impact are, however, significantly suppressed with respect to random codes with the same number of codons in the coding sets. The binary block structure ensuring the robustness of coding for amino acids with high/low hydrophobicity and large/small molecular volumes is also displayed in the distribution of the more stable patterns. Thus, the intrinsic order in the genetic code and the evolutionary selection are quite evident.

The seeming resemblance between the standard code and a partially optimized random code (Novozhilov et al., 2007) needs further investigations. It is likely that the local stability of two codes is distinctly different. In addition, the appearance of approximate binary block structure in a random code after optimization should be checked. Although the search of global minimum in terms of stability function is not rigorously defined, intuitively, we would place the standard code somewhere at the vicinity of one standard deviation near the global minimum in the rugged stability landscape.

The various binary blocks in the standard code may be associated with the different physicochemical characteristics of encoded amino acids. The choice of physicochemical characteristic affects the results of stability analysis and is dictated by the particular problem. For example, blocks NSN-NWN code for amino acids with preferably small/large side-chain volumes. The effects related to side-chain volume are thought to play the important role at the earlier stage of molecular evolution. Evidently, the choice of molecular volume would be natural in the study of possible relationship between these effects and the stability of the genetic code. As amino acids are simultaneously characterized by the multifarious parameters (Xia and Li, 1998), generally, the stability function should depend on several (properly weighted) physicochemical characteristics. The most robust coding for amino acids with several physicochemical characteristics is a typical optimization problem, where the better coding for the overall set may be attained at the expense of partial worsening of coding for particular components.

The more robust modifications of the standard code may induce the development of the alternative variants of the genetic code. The resulting code ought, however, strongly depend on the existing translation mechanisms. Therefore, certain structural features of the standard code should be universal (cf., e.g., the relationship between redundancy over the third codon position and the wobble rules). These restrictions again lead to an optimization problem with possible improvement and worsening in the sets of the different parameters. Table 8 contains the general stability characteristics for the vertebrate mitochondrial code. The comparison with Table 1 for the standard code shows that slightly more robust coding for hydrophobic/hydrophilic amino acids for the mitochondrial code is attained at the expense of the less robust coding for the side-chain volumes and overall characteristics. The weakening of overall coding robustness in the vertebrate mitochondrial code cannot be referred exclusively to the larger number of stop-codons in this code (four versus three in the standard code), which commonly deteriorate the stability of the genetic code (cf. the data in the scheme without stop-codons in Tables 1 and 8).

The error-minimization theory remains likely the most well-posed and verified approach to the problem of the genetic code. The theory permits to assess quantitatively the impact of translation errors and point mutations and to compare the coding robustness of various codes. The absence of constraints related to the critical diversity or complexity makes the minimization problem incomplete and does not allow one to pursue the further steps of potential evolutionary modifications related to more robust coding. The lack of understanding of molecular mechanisms of selection behind the more robust coding also hampers the further progress in this field. The detailed analysis of the available variants of the genetic code might shed additional light on these mechanisms. The other abilities may be related to the study of potential connections between more robust coding and the replacements of amino acids.

The analysis of the local stability of the genetic code proves that the standard code turns out to be relatively labile and contains rich choice of plausible evolutionary variants in its close vicinity. Their systematic investigation may help to find some clues in the fundamental problem of the genetic code and molecular evolution.

**Acknowledgements**

**Table 7**

Relationship between variants of the genetic code and more robust coding

| Codon | Variants of coding | Normalized variations | | |
|---|---|---|---|---|
| | | $\Delta\varphi_H$ | $\Delta\varphi_V$ | $\Delta\varphi_S$ |
| | One-point | Delins | | |
| UAA | Ter → Q | −1.08 | −2.22 | −1.77 |
| | Ter → Y | +0.09 | −1.39 | −0.58 |
| UAG | Ter → Q | −1.38 | −3.07 | −2.37 |
| | Ter → L | −0.47 | −3.03 | −1.73 |
| UGA | Ter → W | −0.64 | −0.23 | −0.54 |
| | Ter → C | −1.27 | −2.11 | −1.85 |
| AUA | I → M | −0.05 | −0.02 | −0.04 |
| AAA | K → N | −0.04 | −0.30 | −0.16 |
| AGA | R → S | −0.26 | −0.88 | −0.58 |
| | R → G | −0.01 | −0.34 | −0.17 |
| AGG | R → S | −0.55 | −0.18 | −0.46 |
| | R → G | −0.42 | +0.59 | −0.01 |
| CGN | Not identified | | | |
| | Possible predictions: | | | |
| | R → S | −0.21 | +0.46 | +0.07 |
| | R → H | −0.25 | −0.22 | −0.27 |
| | R → Q | −0.30 | −0.33 | −0.13 |
| | | Swaps | | |
| UAA | Ter ↔ Q | −0.39 | +0.95 | −0.02 |
| UAG | Ter ↔ Q | −0.58 | −0.02 | −0.52 |
| AAA | K ↔ N | −0.03 | −0.06 | −0.05 |
| AGA | R ↔ S | +0.14 | −0.07 | +0.10 |
| UCA | S ↔ Ter | +0.13 | −0.89 | −0.20 |
| AGA | R ↔ Ter | −0.13 | −1.13 | −0.49 |
| | Two-point | Delins | | |
| UAG | Ter → A | −1.48 | −2.07 | −1.98 |
| CUG | L → S | +1.22 | +1.26 | +1.42 |
| CUN | L → T | +1.10 | +0.77 | +1.11 |

The normalized variations correspond to the scheme with stop-codons and biased conditional probability.

**Table 8**

Characteristics of negative normalized variations of stability function for modifications of the vertebrate mitochondrial code

A. The unbiased conditional probability

| Operation | Total | Fraction with $\Delta\varphi < 0$ | | | Relative fraction with $\Delta\varphi < -1$ | | | Stop-codons |
|---|---|---|---|---|---|---|---|---|
| | | $\varphi_H$ | $\varphi_V$ | $\varphi_S$ | $\varphi_H$ | $\varphi_V$ | $\varphi_S$ | |
| Delin | 444 | 0.41 | 0.40 | 0.37 | 0.22 | 0.31 | 0.32 | Yes |
| Delin | 380 | 0.38 | 0.36 | 0.33 | 0.07 | 0.16 | 0.13 | No |
| Swap | 222 | 0.27 | 0.29 | 0.27 | 0.00 | 0.28 | 0.10 | Yes |
| Swap | 190 | 0.24 | 0.22 | 0.23 | 0.00 | 0.00 | 0.00 | No |

B. The biased conditional probability

| Operation | Total | Fraction with $\Delta\varphi < 0$ | | | Relative fraction with $\Delta\varphi < -1$ | | | Stop-codons |
|---|---|---|---|---|---|---|---|---|
| | | $\varphi_H$ | $\varphi_V$ | $\varphi_S$ | $\varphi_H$ | $\varphi_V$ | $\varphi_S$ | |
| Delin | 444 | 0.31 | 0.35 | 0.32 | 0.14 | 0.25 | 0.30 | Yes |
| Delin | 380 | 0.27 | 0.32 | 0.25 | 0.00 | 0.12 | 0.04 | No |
| Swap | 222 | 0.23 | 0.28 | 0.22 | 0.00 | 0.10 | 0.08 | Yes |
| Swap | 190 | 0.16 | 0.15 | 0.11 | 0.00 | 0.00 | 0.00 | No |

For random codes, the expected fractions for $\Delta\varphi < 0$ and $\Delta\varphi < -1$ are 0.5 and 0.32, respectively. Relative fraction with $\Delta\varphi < -1$ is defined with respect to the number of variations with $\Delta\varphi < 0$.